\documentclass[useAMS,usenatbib]{mn2e}

\input{epsf}
\usepackage{epsfig}
\usepackage{amssymb}
\usepackage{times}

\def\be{\begin{equation}}
\def\ee{\end{equation}}
\def\bc{\begin{center}}
\def\ec{\end{center}}
\def\beq{\begin{eqnarray}}
\def\eeq{\end{eqnarray}}

\def\msun{{\rm M_{\odot}}}
\def\d{{\rm d}}

\def\Te{T_{\rm e}}
\def\phip{\phi_{\rm p}}
\def\phis{\phi_{\rm s}}
\def\alphap{\alpha_{\rm p}}
\def\alphas{\alpha_{\rm s}}

\def\rg{r_{\rm S}} % Schwarzschild radius
\def\betaeq{\beta_{\rm eq}}
\def\Dop{\delta}

\def\source{SAX J1808.4$-$3658}
\def\mumin{\mu_{\rm min}}
\def\mumax{\mu_{\rm max}}
\def\phiobs{\phi_{\rm obs}}
\def\ani{h}

\title[Pulse profiles of  millisecond pulsars]
{Pulse profiles  of   millisecond pulsars and their Fourier amplitudes}

\author[J. Poutanen and A. M. Beloborodov]
{Juri~Poutanen$^{1}$\thanks{E-mail:
juri.poutanen@oulu.fi (JP); amb@phys.columbia.edu (AMB)}
and  Andrei M. Beloborodov$^{2,3}$\footnotemark[1] \\
$^{1}$ Astronomy Division, P.O. Box 3000, FIN-90014 University of Oulu,
Finland  \\
$^{2}$ Physics Department and Columbia Astrophysics Laboratory,
 Columbia University, 538 West 120th Street, New York, NY 10027, USA \\
$^{3}$  Astro Space Center, Lebedev Physical Institute,
           Profsojuznaja 84/32, 117810 Moscow, Russia}

\date{Accepted 2006 September 20. Received 2006 August 30}
\pagerange{836--844} 
\volume{373}
\pubyear{2006}
\begin{document}

\maketitle

\label{firstpage}

\begin{abstract}
Approximate analytical formulae are derived for the pulse profile produced by small hot 
spots on a  rapidly rotating neutron star.
Its Fourier amplitudes and phases are calculated. 
The proposed formalism takes into account gravitational bending of light, 
Doppler effect, anisotropy of emission, and time delays. 
Its accuracy is checked with exact numerical calculations. 
\end{abstract}


\section{Introduction}

During the past decade, coherent (or nearly coherent) oscillations
in the $\sim$200--600 Hz frequency range were discovered
in the  light curves of a number
of neutron stars in low-mass X-ray binaries observed by
the {\it Rossi X-ray Timing Explorer (RXTE)}.
In 13 sources,
these oscillations were discovered during X-ray bursts
\citep[see][ for a review]{sb06}, giving the name {\it nuclear-powered
millisecond pulsars} to this class of objects.
Seven transient sources, {\it accretion-powered millisecond pulsars},
showed coherent pulsations in the persistent flux during the outbursts
lasting   a few weeks \citep[see reviews by ][]{p06,w05}.
In all these millisecond pulsars (hereafter MSP), the observed emission is dominated
by bright spots on the neutron star surface.
Such bright spots are created either by a thermonuclear explosion observed
as an X-ray burst or by an accretion flow channelled towards a magnetic
pole. The pulse profile produced by the spot
carries information about the position of the spot, its size, spectrum
and angular distribution of its emission and the gravitational field of
the star.

The X-ray pulse profiles observed from MSP
are almost sinusoidal. In \source\ \citep{gdb02,pg03}
and XTE~J1814-338 \citep{s03},  there is a noticeable
skewing of the profile which increases with photon energy.
The skewness as well as the observed
soft time lags probably result from Doppler boost
of anisotropically emitted radiation \citep{gdb02,pg03}.
Pulse profiles of \source\  have been measured with high accuracy at 
different photon energies and well fitted by a theoretical model  \citep{pg03}.
This gave constraints   on the neutron star   radius  
$8<R< 12$ km (assuming mass of 1.4--1.6$\msun$) and
on the inclination  of the spin axis to the line of sight $i\gtrsim60\degr$.

The poor photon statistics available for MSP normally
does not allow one to study in detail the shapes of their light curves.
Therefore,   data analysis is often limited to the amplitude and
phase of the fundamental Fourier harmonic, and sometimes higher
harmonics may be analyzed.
The Fourier technique helps use the available data to constrain
the neutron star parameters
\citep*[see e.g.][]{ml98,wml01,moc02,gp05}.
A number of recent papers were devoted to numerical calculations
of the amplitude of pulsations created by a spot (or two antipodal
spots) for different radiation patterns, rotational velocities,
inclinations,
spot positions and their sizes \citep*[see e.g.][]{wml01,moc02}.
Such calculations involve light-bending and Doppler effects,
which made the problem complicated and required numerical calculations.
Given the large number of parameters, the numerical approach makes it
difficult to understand the dependence of results on parameters and
interpret the data.

The purpose of the present paper is to develop an approximate
analytical description of the problem.
Using the simple formalism  of  \citet[][ hereafter B02]{b02}
for light bending, we derive analytical formulae for oscillation amplitudes
which demonstrate how the observables depend on parameters of the problem.

The plan of the paper is as follows. In Section \ref{sec:model},
we introduce our notations and summarize the exact numerical method of
the light-curve calculation.
In Section \ref{sec:anal} we discuss the approximate description of light
bending, introduce an approximate formula for time-delay effects, 
describe possible classes of MSP, and finally 
derive approximate analytical formulae for the pulse profiles and
the corresponding  amplitudes and phases of the Fourier series. 
Using our analytical formalism  we compare the approximate Fourier amplitudes to the exact results
and  investigate the effects of anisotropy  of the emission pattern and the Doppler effect 
on the pulse profile  in Section \ref{sec:results}.

%#########################################################################

\section{Pulse from a spot on a spinning star}
\label{sec:model}

\begin{figure}
\centerline{\epsfig{file=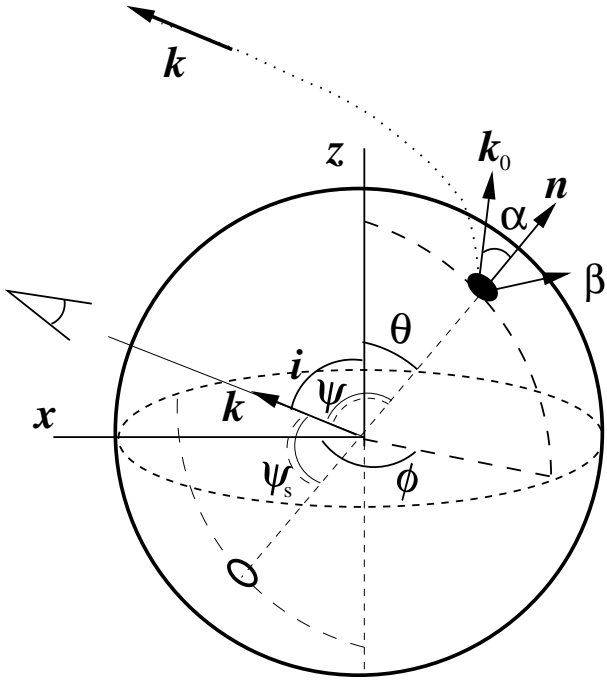,width=8.0cm}}
\caption{Geometry of the problem. Dotted curve shows the
photon trajectory.
\label{fig:geom}}
\end{figure}

Consider a small spot on the star surface. Its area measured in the
corotating frame is $dS^\prime$, and its instantaneous position in
the fixed lab frame is described by the unit vector $\bmath{n}$
that points to the spot from the star center (see Fig.~\ref{fig:geom}).
The angle between $\bmath{n}$ and the line of sight is denoted by $\psi$.
We are interested in photons emitted by the spot that propagate
along our line of sight at large distances from the star (where
gravitational bending becomes negligible). We denote the unit vector
along the line of sight by $\bmath{k}$, so that
\be
  \cos\psi=\bmath{k}\cdot \bmath{n}.
\ee
As the star rotates, $\bmath{k}\cdot\bmath{n}$ varies periodically,
\be \label{eq:psi}
\cos\psi=\cos i\ \cos\theta+\sin i\ \sin \theta\ \cos\phi,
\ee
where $i$ is the inclination angle of the spin axis to the line of sight, $\theta$ is the 
spot colatitude  and
$\phi=2\pi\nu t$ is the rotational phase of the pulsar;
$\nu=P^{-1}$ is the pulsar frequency, and $t=0$ is chosen when the spot
is closest to the observer.

%%%%%%%%%%%%%%%%%%%%%%%%%%%%%%%%%%%%%%%%%%%%%
%%%%%%%%%%%%%%%%%%%%%%%%%%%%%%%%%%%%%%%%%%%%%

Angle $\psi$ measures the apparent inclination of the spot to the line of
sight, which is different from the true inclination because of
the light bending effect.
% The initial angle of the emitted photon with respect to the normal
% ${\mathbf n}$ differs from $\psi$ because of the light bending.
We denote the initial direction of the emitted photon by $\bmath{k}_0$
and the true emission angle by $\alpha$, so that
\be
 \cos\alpha=\bmath{k}_0 \cdot \bmath{n}.
\ee

Emission angle in the corotating frame is denoted by
$\alpha'$.  It differs from $\alpha$ because of relativistic
aberration (see derivation in the Appendix)
\be \label{eq:aberr}
\cos\alpha' =   \Dop \ \cos\alpha ,
\ee
where $\Dop =1/\gamma(1-\beta\cos\xi)$
is the Doppler factor.
Here $\gamma=(1-\beta^2)^{-1/2}$ and $\beta=v/c$ is the spot velocity,
\be\label{eq:beta}
\beta=\frac{2\pi R}{c} \frac{\nu}{\sqrt{1-u}} \sin\theta =\betaeq \sin\theta,
\ee
$\betaeq$ is the velocity at the equator and $\xi$ is the angle  between
the spot velocity and $\bmath{k}_0$.  Here $u\equiv\rg/R$, 
$\rg=2GM/c^2$ is the Schwarzschild radius; $M$ and $R$ are mass and
radius of the star.
The pulsar frequency has been corrected for the redshift $\sqrt{1-u}$.
% Using equation~(\ref{eq:k0}) it is easy to show that
One can show that $\xi$ is related to $\alpha$, $\psi$, $i$ and $\phi$ by
(see Appendix),
\be \label{eq:cosxi}
\cos\xi=
% \frac{\bbeta}{\beta} \cdot \bmath{k}_0
% =\frac{\sin\alpha}{\sin\psi} \frac{\bbeta}{\beta} \cdot \bmath{k}=
- \frac{\sin\alpha}{\sin\psi}\sin i\ \sin\phi\  .
\ee

%%%%%%%%%%%%%%%%%%%%%%%%%%%%%%%%%%%%%%%%%%%%%
\begin{figure}
\centerline{\epsfig{file=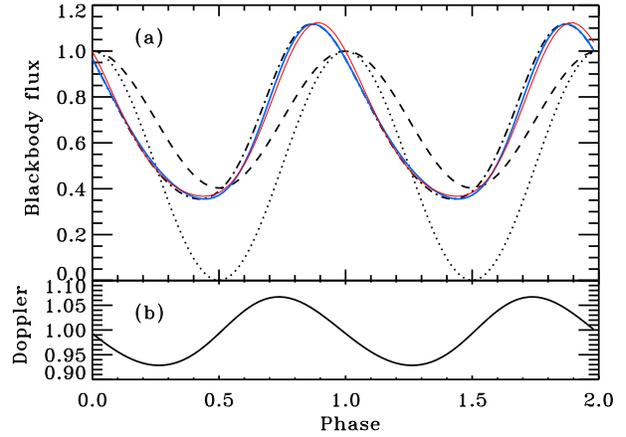,width=8.0cm}}
\caption{
(a) The bolometric black body flux  as a function of the observed phase.
Dotted curve is for a slowly rotating star ignoring all relativistic effects.
The pulse profile where 
gravitational light bending is accounted for is shown by the dashed curve.
Dot-dashed curve gives the profile modified by the Doppler boost
and aberration for a neutron star rotational frequency $\nu=600$ Hz.
Solid curve accounts also for the time delay.
Thin solid curve is a pulse profile produced using the approximate 
Fourier amplitudes derived in Sect. \ref{sec:anal}. 
(b)  Doppler factor $\Dop$ as a function of the observed phase
for  $\nu=600$ Hz. We take $i=\theta=45\degr$, $R=2.5\rg$, $M=1.4\msun$ in
this example.
\label{fig:relat_effects}}
\end{figure}
%%%%%%%%%%%%%%%%%%%%%%%%%%%%%%%%%%%%%%%%%%%%%

For power-law spectra (observed for example 
in accretion-powered MSP),
we assume that the energy and angular dependencies of the spectrum
emitted  by the spot may be separated as
\be \label{eq:Isepar}
I_{E'}(\alpha') = I_0 (1+ \ani\cos\alpha')E'^{-(\Gamma-1)} ,
\ee
where $\ani$ does not depend on photon energy $E$.
If the power-law spectrum
is produced by thermal Comptonization by electrons of temperature
$T_{\rm e}$, this condition would be satisfied if the maximum Doppler shift
$\Delta \delta \sim 2\pi \nu R/c \ \sin i\ \sin\theta$ is smaller than
the typical relative energy change in a single scattering
$\Delta E/E \sim 4k\Te/m_{\rm e} c^2$, which translates to
$ (\nu/600\ \mbox{Hz}) \sin i\ \sin\theta < k\Te/16\ \mbox{keV}$ \citep{vp04}.
Even for X-ray burst spectra one  expects that $\ani$ varies slowly with
energy, so that equation (\ref{eq:Isepar}) still can be used.

The observed spectral flux at a distance $D$ from the star is then given by
(see derivation in Appendix),
\be\label{eq:fluxplaw}
 F_{E}= (1-u)^{\Gamma/2}\ \Dop^{\Gamma+3} I'_E(\alpha')
\cos\alpha\ \frac{\d\cos\alpha}{\d\cos\psi}\ \frac{\d S'}{D^2}.
\ee
Expression for the bolometric flux
may be obtained as a special case of equation~(\ref{eq:fluxplaw}) by setting
$\Gamma=2$,
\be \label{eq:fluxbolo}
 F= (1-u)\ \Dop^5 \
I'(\alpha') \ \cos\alpha\ \frac{\d\cos\alpha}{\d\cos\psi}\ \frac{\d S'}{D^2} .
\ee
These equations take into account
the special relativistic effects (Doppler boost, relativistic aberration)
as well as general relativistic effects (gravitational redshift and light
bending in Schwarzschild geometry).

For further analysis we use pulse profiles normalized to 
$F_0= I_0 E^{-(\Gamma-1)} (1-u)^{(\Gamma+2)/2}\ \d S' / D^2$: 
\be \label{eq:normflux_exact}
%\bar{F}(\phi)\equiv \frac{
F(\phi) = 
%}{F_0} =
\Dop^{\Gamma+3} (1+\ani \Dop\cos\alpha ) \cos\alpha
\ \frac{1}{1-u}\ \frac{\d\cos\alpha}{\d\cos\psi}.
\ee
The flux is zero if $\cos\alpha<0$.
%The flux is zero if  $\psi>\psi_{\max}$.
For the  antipodal  spot, we  substitute
$\theta \rightarrow \pi-\theta$ and $\phi\rightarrow \pi+\phi$.

%%%%%%%%%%%%%%%%%%%%%%%%%%%%%%%%%%%%%%%%%%%%%%%%%%
\begin{figure}
\centerline{\epsfig{file=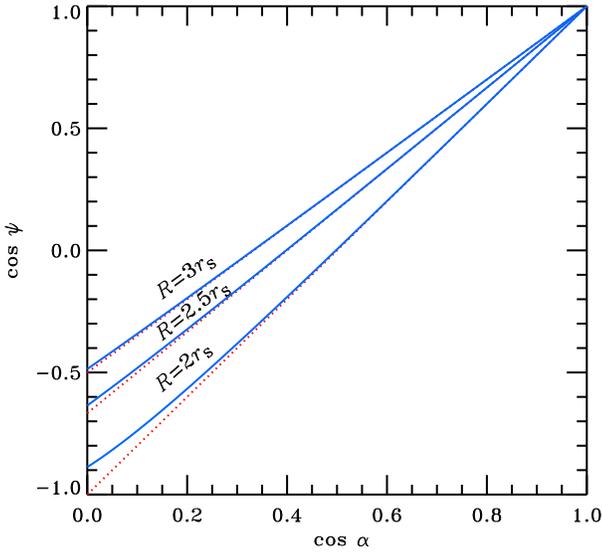,width=8.0cm}}
\caption{
Light bending in Schwarzschild metric. The solid curves are the exact
results using equation (\ref{eq:bend}) and the
dotted lines is the approximation (\ref{eq:cosbend}).
\label{fig:bending}}
\end{figure}
%%%%%%%%%%%%%%%%%%%%%%%%%%%%%%%%%%%%%%%%%%%%%%%%%%

Expression (\ref{eq:normflux_exact}) does not account for time delays resulting from
different paths travelled by photons emitted at different phases $\phi$.
The delays become significant only for very fast-rotating pulsars.
In Schwarzschild metric the maximum time delay for a neutron star
of $M=1.4\msun$ is $\Delta t\sim 7\times 10^{-2}$ ms (almost independent
of compactness of the star $M/R$). This gives at most 
a 5 per cent correction to the arrival phase 
for a rotational period $P=1.5$~ms.
The flux at observed phase $\phiobs$ is
$\bar{F}(\phiobs)=F(\phiobs-\Delta\phi)$ with
phase delay  $\Delta \phi=2\pi\nu\Delta t$
computed using (\ref{eq:delay}) and (\ref{eq:deltaphi}) from the Appendix.
The effect of the photon arrival time contraction (or stretching) on
the observed flux is already accounted for by one of the Doppler factors.
The effects of gravitational bending, Doppler boost, and time delays 
on the pulse profile are shown in Fig. \ref{fig:relat_effects}.

Fourier series of the pulse profile is given by
\be \label{eq:genfor}
\bar{F}(\phiobs) = A_0+ \sum_{n=1} [ A_n \cos (n\phiobs) + B_n \sin (n\phiobs)] ,
\ee
where
\beq \label{eq:fourier_coeff}
A_0&=& \frac{1}{2\pi} \int_0^{2\pi} \bar{F}(\phiobs)\ \d \phiobs , \nonumber \\
A_n&=& \frac{1}{\pi} \int_0^{2\pi}  \bar{F}(\phiobs) \cos (n\phiobs)\ \d \phiobs , \ n\ge 1\\
B_n&=& \frac{1}{\pi} \int_0^{2\pi}  \bar{F}(\phiobs) \sin (n\phiobs)\ \d \phiobs, \ n\ge 1 . \nonumber
\eeq
An alternative form of Fourier series is written in terms of amplitudes $c_n$ and 
phases $\Delta\phi_n$
\be  \label{eq:fourier}
\bar{F}(\phiobs) = \sum_{n=0}  C_n \cos[ n (\phiobs+\Delta\phi_n)] ,
\ee
where
\be \label{eq:cn}
C_n=\sqrt{A_n^2+B_n^2}, \quad \tan (n\Delta\phi_n) =-B_n/A_n.
\ee
The exact Fourier series are calculated numerically.

\section{Analytical approximation}
\label{sec:anal}

\subsection{Light bending and time delay}

The expressions for the observed flux (\ref{eq:fluxplaw}) and (\ref{eq:fluxbolo}) 
can be significantly simplified if one uses analytical formula
for light bending and time delays. This will allow us to also obtain
analytical expressions for the pulsation amplitude and Fourier harmonics
in Section \ref{sec:four}.

\citet{b02} showed that the relation
\be\label{eq:cosbend}
\cos\alpha\approx u + (1-u) \cos\psi
\ee
describes light bending with high accuracy \citep[see also][ for discussion 
of approximations]{zsp95,ll95}.
The accuracy of equation (\ref{eq:cosbend}) is shown in Fig.~\ref{fig:bending}
(see also B02).
For a star with $R=2\rg$, the accuracy is better than 10 per cent,
while  for $R=3\rg$ the error does not exceed 3 per cent.
%In this approximation $\d\cos\alpha/\d\cos\psi = 1-u$ and 
The spot is visible to the observer when 
\be \label{eq:visib}
\cos\psi>\cos\psi_{\max} = -u/(1-u).
\ee

%%%%%%%%%%%%%%%%%%%%%%%%%%%%%%%%%%%%%%%%%%%%%%%%%%
\begin{figure}
\centerline{\epsfig{file=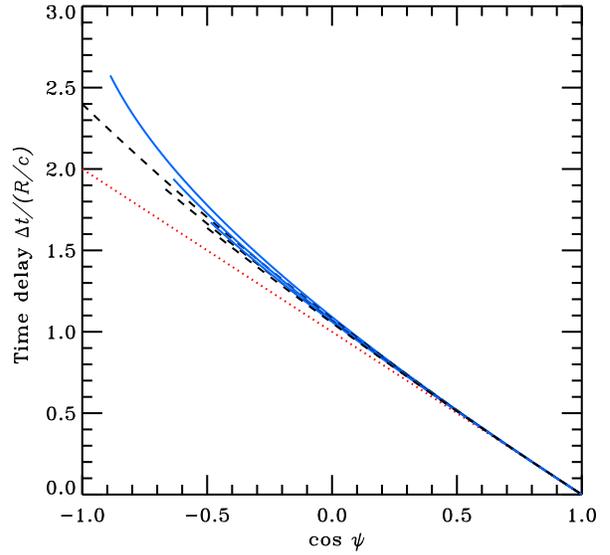,width=8.0cm}}
\caption{
Time delays in Schwarzschild metric. The solid curves are the exact
results using equation (\ref{eq:delay}) for $R=2, 2.5$ and $3\rg$ (from top to bottom).
The dashed curves show the results of an approximate formula
(\ref{eq:delayapp}), and the dotted line is the approximation (\ref{eq:delaysim}).
\label{fig:timedelay}}
\end{figure}
%%%%%%%%%%%%%%%%%%%%%%%%%%%%%%%%%%%%%%%%%%%%%%%%%%

%%%%%%%%%%%%%%%%%%%%%%%%%%%%%%%%%%%%%%%%%%%%%%%%%%%%%%%
\begin{figure*}
\centerline{\epsfig{file=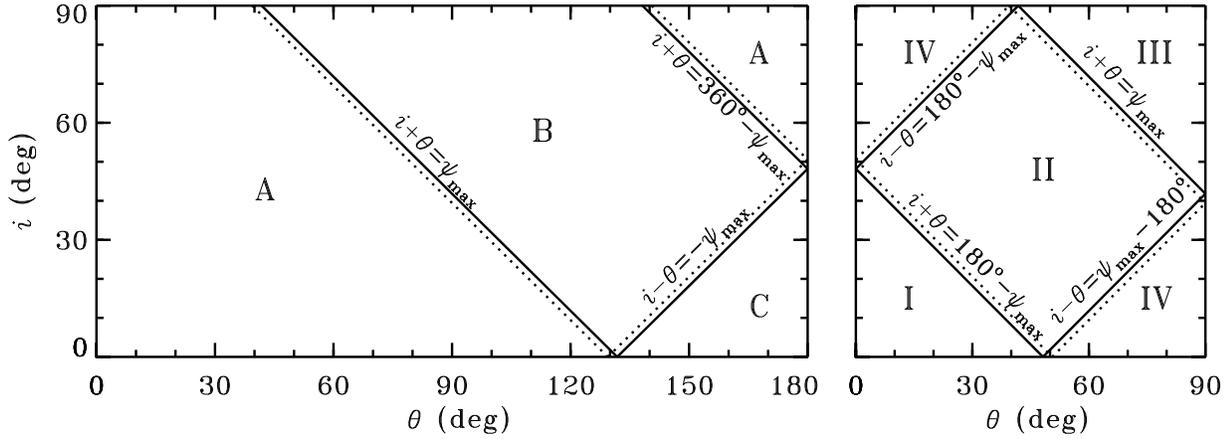,width=16cm}}
\caption{Classes of visibility for hot spots on neutron star surface 
at the inclination $i$-- spot colatitude $\theta$ plane.
%Location of different classes on  for one spot (left panel) and two antipodal spots (right panel).
The classes are determined by the relative positions of
$\mumin=\cos(i+\theta)$, $\mumax=\cos(i-\theta)$, $\cos\psi_{\max}$ and
$\kappa\equiv|\cos\psi_{\max} |$.
Left panel: classes for one spot. 
In class A  defined by $\mumin>-\kappa$ 
the spot is always visible.
In class B defined by $\mumin<-\kappa<\mumax$
the   spot is not visible for a fraction of period.
In class C defined by
$\mumax<-\kappa$ the spot is never visible.
%The angle $\xi$ is defined so that $\cos\xi=r_g/(R-r_g)$,
%for $R=2.5r_g$.
Right panel: classification for two antipodal spots as given in B02.
In class~I, $\mumin>\kappa$,
the primary spot is visible all the time and the antipodal spot is never seen.
In class~II, $-\kappa<\mumin<\kappa<\mumax$,
the primary spot is seen all the time and the antipodal spot also appears for some time.
In class~III, $\mumin<-\kappa$, the primary spot is not visible for a
fraction of period (and then only the antipodal spot is seen).
In class~IV, $-\kappa<\mumin,\mumax<\kappa$, both spots are seen at any
time. Dotted and solid lines correspond to the exact $\psi_{\max}=129\fdg4$
and approximate $\cos\psi_{\max}=-u/(1-u)$ (giving $\psi_{\max}=131\fdg8$),
for $u=0.4$  (i.e. $R=2.5\rg$).
\label{fig:class}}
\end{figure*}
%%%%%%%%%%%%%%%%%%%%%%%%%%%%%%%%%%%%%%%%%%%%%%%%%%%%%%%

The exact expression for the time delays (\ref{eq:delay}) may also be
approximated by a simple formula.
Expanding it in Taylor series with $1-\cos\alpha$ as a small parameter and
using also the expansion
\be
\frac{1-\cos\alpha}{1-u}=y \left( 1+ \frac{u^2}{112} y^2 \right)
\ee
(obtained from eq.~[2] in B02), where $y=1-\cos\psi$, we get
\be \label{eq:delayapp}
\Delta t =   y \left\{ 1 + \frac{uy}{8}\left[ 1+y(1/3 -u/14)\right] \right\}\ R/c .
\ee
Keeping only the leading term, we find
\be\label{eq:delaysim}
\Delta t  = y \ R/c .
\ee
This approximation
is  better than 20 per cent  accurate for most emission angles and
compactnesses (Fig.~\ref{fig:timedelay}). Only in the extreme case of $R=2\rg$ for
large bending angles (and large delays), the error increases to 35 per cent.
Since the time delays themselves produce a small effect, equation
(\ref{eq:delaysim}) is sufficient in most calculations.

%##########################################################################
\subsection{Pulsar  visibility classes}

Analytical light bending formula (\ref{eq:cosbend}) allows one to 
introduce a simple classification of the light curves according to the relative
positions of $\mumin\equiv \cos(i+\theta)$, $\mumax\equiv \cos(i-\theta)$,
$\cos\psi_{\max}$ and $\kappa\equiv|\cos\psi_{\max}|$.
For a single spot,   three classes exist  (see left panel Fig.~\ref{fig:class}).
In class A, defined by  $\mumin>-\kappa$, the spot is always visible.
When $\kappa<-\mumax$ (class C), the spot is always invisible.
For $\mumin<-\kappa<\mumax$ (class B), the spot is visible during the pulsar phases
$|\phi |< \phip$, where
\be \label{eq:phip}
\cos\phip =(\cos\psi_{\max}  - \cos i\cos\theta)/\sin i \sin \theta =-Q/U,
\ee
and we defined
\beq
U&=&(1-u) \sin i\ \sin \theta , \nonumber \\
Q&=& u+(1-u) \cos i \ \cos\theta .
\eeq

Pulsars with two antipodal spots  are divided into four classes 
(shown in the right panel of Fig.~\ref{fig:class}, see also B02).
In class I, corresponding to $\mumin> \kappa$
only the primary spot is  visible all the time.
In class II, $-\kappa<\mumin<  \kappa <\mumax$,
the primary spot is always visible, while 
the antipodal  secondary spot appears during phases
$\phis < \phi < 2\pi-\phis$, where
\be \label{eq:phis}
\cos\phis = -\frac{\cos\psi_{\max}  + \cos i\cos\theta}{\sin i \sin \theta} =\frac{2u-Q}{U}.
\ee
The primary spot disappears for a fraction of the period
in   class   III, $\mumin<-\kappa$,  and   then   only the   antipodal   spot      is   seen.
And finally, in class IV ($-\kappa<\mumin,
\mumax<  \kappa $), both spots are seen all  the  time.

\subsection{Fourier series}
\label{sec:four}

Our aim is to obtain simple analytical expressions for
the Fourier amplitudes and phases characterizing the pulse profile.
We can simplify expression (\ref{eq:normflux_exact}) by
using approximation  to the light bending  formula (\ref{eq:cosbend}):
\be\label{eq:fluxgen}
F(\phi) =  \Dop^{\Gamma+3} (1+ \ani \Dop\cos\alpha ) \cos\alpha    ,
\ee
where now
\be
\cos\alpha=
%u+(1-u)[\cos i\cos\theta+\sin i\sin \theta\cos\phi] =
Q+U \cos\phi .
\ee
%is a function of $\cos\phi$ only.
Doppler factor $\Dop$ depends on $\betaeq$. 
In the leading order of  $\betaeq\ll 1$ this dependence is given by 
\be \label{eq:dopexa}
\Dop \approx 1- \betaeq \sin  i\   \sin\theta   \frac{\sin\alpha}{\sin\psi} \sin\phi .
\ee
We further approximate
$\sin\alpha/\sin\psi\approx \sqrt{1-u}$  which becomes exact at $\alpha\ll1$ 
(cf. the cosine relation [\ref{eq:cosbend}]). This gives
\be \label{eq:dopapp}
\Dop \approx 1- T \sin\phi,
\ee
where 
\be \label{eq:t}
T \equiv \betaeq  \sqrt{1-u}\ \sin i\  \sin  \theta \ll 1. 
\ee 
Hereafter, we keep linear terms in $T$ and neglect 
higher order terms. Then, substituting $\Dop$ 
into expression (\ref{eq:fluxgen}), we obtain
the flux (as a function of $\phi$):
\be
F(\phi) = a_0 + \sum_{n=1}^{3} [  a_n \cos (n\phi) + b_n \sin (n\phi)] ,
\ee
with non-zero coefficients 
\beq 
\label{eq:a0} a_0 &=& Q+ \ani (Q^2+U^2/2) ,  \\ 
\label{eq:a1} a_1 &=& (1+2 \ani  Q) U ,   \\ 
\label{eq:b1} b_1 &=& - \left[ Q(3+\Gamma)+ \ani \left( Q^2+\frac{U^2}{4}\right) (4+\Gamma)\right] \ T ,   \\
\label{eq:a2} a_2 &=& \ani\  U^2/2 ,  \\
\label{eq:b2} b_2 &=& - [(1+2\ani Q) (4+\Gamma)-1] \ \frac{TU}{2} ,   \\
\label{eq:b3} b_3 &=& - \frac{4+\Gamma}{4} \ \ani\ T U^2 .  
\eeq
Similarly to equation (\ref{eq:cn}), we can define Fourier coefficients $c_n$ and 
corresponding phase lags. 
Coefficients (\ref{eq:a0})--(\ref{eq:b3}) are good approximations to the exact Fourier coefficients
$A_n, B_n$ of the flux $\bar{F}(\phiobs)$
as long as the spot is visible at all phases (through the entire 
period of rotation). 
Alternatively, the pulse profile can be written   as a cosine series
\beq
F (\phi) 
&=&  Q+ \ani (Q^2+U^2/2)  \nonumber\\ 
&+& \frac{U(1+2 \ani Q)}{\cos\zeta_1} \cos(\phi+\zeta_1)     \\
&+& \frac{\ani U^2/2 }{\cos2\zeta_2} \cos[2(\phi+\zeta_2)] 
+ \frac{4+\Gamma}{4} \ani T U^2 \cos[3(\phi+\pi/6)] , \nonumber
\eeq
where
\beq \label{eq:zeta1}
\tan \zeta_1&=&\frac{T}{U} \ 
\frac{(3+\Gamma)Q+(4+\Gamma) \ani (Q^2+U^2/4)}{1+2 \ani Q} , \\
\label{eq:zeta2} 
\tan2\zeta_2&=&\frac{T}{U} \  \frac{(4+\Gamma)(1+ 2\ani Q)-1 }{\ani} .
\eeq
The ratio of the harmonic to the fundamental  
\be
\frac{c_2}{c_1}=\frac{\ani U}{2(1+2 \ani Q)} \frac{\cos\zeta_1}{\cos2\zeta_2} 
\ee
grows with the anisotropy parameter $\ani$ and $\sin i\sin\theta$.

If the spot disappears from the visibility zone 
for part of the period (classes B, II, and III),  the 
Fourier series becomes infinite.
The Fourier coefficients of the light curve
can be computed from the coefficients (\ref{eq:a0})--(\ref{eq:b3}):
\beq \label{eq:four_ecl}
a'_0 &=& \frac{\phip }{\pi} \sum_{k=0}^3 a_k s_k(\phip) , \nonumber \\
a'_n &=& \frac{\phip }{\pi} \sum_{k=0}^3 a_k  [ s_{n-k}(\phip) + s_{n+k}(\phip) ] , \\
b'_n &=& \frac{\phip }{\pi} \sum_{k=1}^3 b_k   [ s_{n-k}(\phip) - s_{n+k}(\phip) ] , \nonumber
\eeq
where $s_0(\phi)=1$ and $s_n(\phi)=\sin(n\phi)/(n\phi)$.

Including the time delays further modify the expansions.
We are interested in the Fourier coefficients of the function
$\bar{F}(\phiobs)=F(\phiobs-\Delta\phi(\phi))$.
We can calculate the phase delays relative to the photons arriving 
from the star element closest to the observer  (with impact parameter $b=0$)
using equation (\ref{eq:delaysim}):
\be
\Delta \phi(\phi) \approx\Delta \phi(\phiobs)   \approx \Phi - T \cos\phiobs .
\ee
where $\Phi= \betaeq  \sqrt{1-u}(1-\cos i\cos\theta)= (1-Q)  \betaeq /\sqrt{1-u}$.
Keeping only  the first term in Taylor expansion, we arrive at
\beq
\cos n(\phi-\Delta\phi) &=&  \cos n\phi + n \Delta\phi \sin  n\phi , \nonumber \\
\sin n(\phi-\Delta\phi) &=&  \sin n\phi  - n \Delta\phi \cos  n\phi .
\eeq
The Fourier amplitudes for $\bar{F}(\phiobs)$ are found as follows
\beq \label{eq:ab2prime}
%a'_0 &=& a_0   + \frac{T}{2}  b_1  , \\ 
a''_n &=& a_n , \\ % - \Phi n b _n  + \frac{T}{2} \left[  (n-1) b_{n-1} + (n+1) b_{n+1} \right],   \\
b''_n &=& b_n + \Phi n a _n  -\frac{T}{2} \left[  (n-1) a_{n-1} + (n+1) a_{n+1} \right] , \nonumber
\eeq
where we neglected products $Tb_n, \Phi b_n \propto T^2$.
If the spot is invisible for a part of the period, 
one should use coefficients $a'_n, b'_n$ instead of $a_n, b_n$.
For the  antipodal  spot, we  substitute
$\theta \rightarrow \pi-\theta$ and $\phi\rightarrow \pi+\phi$. 
Thus we still can use the expressions (\ref{eq:a0})-(\ref{eq:b3}), 
where $Q$ is replaced by $Q_{\rm s}=u-(1-u) \cos i\ \cos\theta$ and the sign 
of the odd terms $a_1, b_1, b_3$ is changed.

\section{Results}

\label{sec:results}

   We have checked the accuracy of the analytical formalism (Sect.~\ref{sec:four}) 
   by direct comparison with the exact numerical calculation (Sect.~\ref{sec:model}).
The accuracy depends mainly on the compactness of the star $u =\rg/R$ 
   and its rotational frequency $\nu$.
   Part of the error comes from the light-bending approximation (\ref{eq:cosbend});
   this error increases for stars with large $u$ (Fig.~\ref{fig:bending}). Besides,
   we made the approximation $\sin\alpha/\sin\psi\approx (1-u)^{1/2}$ 
   in equation~(\ref{eq:dopexa}) for the Doppler factor $\Dop$.
   This introduces an additional error which becomes noticeable for fast 
   rotators. Errors generally grow at higher $\nu$ because we neglected
   the terms quadratic in $\betaeq$ in all formulae. 
 
\begin{figure*}
\centerline{\epsfig{file=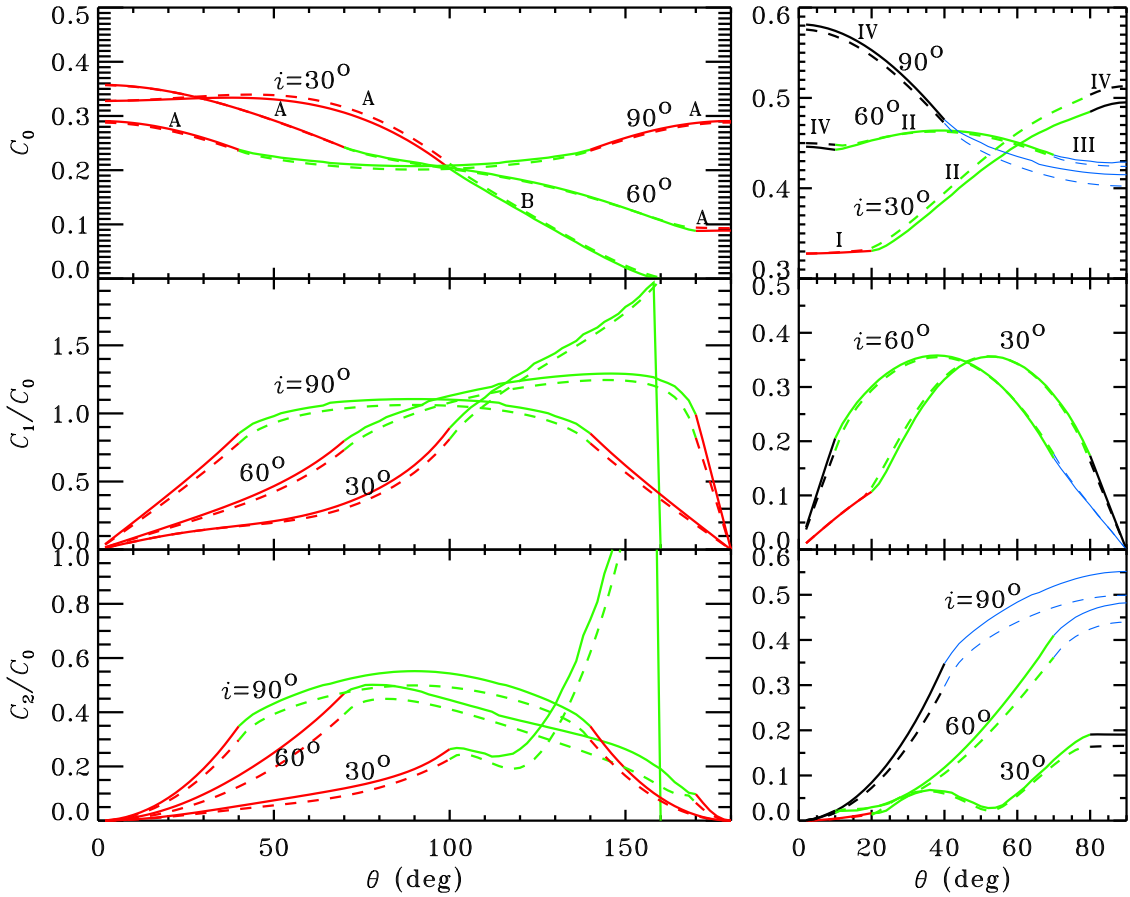,width=15cm}}
\caption{ 
Fourier amplitudes of the pulse profile as functions of the spot colatitude 
for three inclinations $i=30\degr, 60\degr, 90\degr$. 
Left panel: one spot. Right panel: two antipodal spots. Solid curves 
show the exact numerical results and dashed curves -- 
the analytical approximation. 
Inclinations and corresponding visibility classes are marked above the curves.
Parameters are $M=1.4\msun$, $R=2.5\rg$, 
$\Gamma=2$, $\nu=600$ Hz,  $\ani=-0.7$. 
\label{fig:ampl}}
\end{figure*}

As an example, we show   in Figures~\ref{fig:relat_effects} and \ref{fig:ampl} 
the results for a fast and  compact rotator with $\nu=600$~Hz and $u=0.4$ ($R/\rg=2.5$). 
In  this case, the pulse profile reconstructed from the analytical Fourier amplitudes  
is very close to the exact  profile (Fig.~\ref{fig:relat_effects}).
The accuracy of analytical approximation  for amplitudes $C_0$ and $C_1$  
is  better than 3 per cent, while $C_2$ is 15 per cent accurate (see Fig. \ref{fig:ampl}).  
The phases $\zeta_1$ and $\zeta_2$ (eqs.~[\ref{eq:zeta1}] and [\ref{eq:zeta2}], and 
the phases obtained from eqs. [\ref{eq:four_ecl}] and [\ref{eq:ab2prime}]) are accurate within $0.2$ rad.
   For slower rotation, e.g. $\nu=300$~Hz, the error of analytical 
   approximation decreases to $\sim 5$ per cent. Only amplitude $C_3$ 
   (which is smaller than $C_1$ and $C_2$ and more sensitive to the 
   neglected terms in the analytical expansion) has a significant error,
our formulae underestimate $C_3$ by about 50 (20) per cent for $\nu$=600 (300) Hz.
 
    Even for an extremely compact star with $u=0.5$, the analytical 
   $C_1$ and $C_2$ have a good accuracy: they are typically 
   10--20 per cent smaller than the exact values. 
   Only in the cases with one spot at large colatitudes $\theta>120\degr$ and 
   small inclinations $i<60\degr$, the amplitudes are underestimated by a factor 
   of 2, because of extreme gravitational bending.
   
%%%%%%%%%%%%%%%%%%%%%%%%%%%%%%%%%%%%%%%%%%%%%%%%%%%%
%%%%%%%%%%%%%%%%%%%%%%%%%%%%%%%%%%%%%%%%%%%%%%%%%%%%

Using   the analytical formalism one can understand the behaviour 
of Fourier amplitudes and their ratio obtained previously by 
numerical calculations \citep{wml01,moc02}.
  Below we investigate separately the effects of anisotropy 
  and fast rotation.

\subsection{Effects of anisotropy}
\label{sec:ani}

Let us assume a slowly rotating star (i.e. take $\Dop=1$ and $\phiobs=\phi$)
and investigate the effects of anisotropy $\ani\neq 0$ of the source emission
$I(\alpha)=I_0 (1+\ani\cos\alpha)$ on the pulse profile. 
For example, radiation from an optically thin source (slab) is more beamed along the 
stellar surface and described by $\ani<0$ \citep[see e.g.][]{pg03,vp04},
while $\ani\sim2$  characterizes  radiation from
an  optically   thick   electron-scattering dominated atmosphere.

 As will be seen from equations below, anisotropy (i) introduces harmonic
$a_2\propto \ani$, (ii) changes the global structure of the profile
(e.g. two maxima may appear and $\phi=0$ may become a minimum), and 
(iii) leaves the profile symmetric.

 We will use the following expression for the flux from a single  (or primary) spot,
\be
F_{\rm p} (\phi) = \cos\alphap [ 1+\ani\cos\alphap ] ,
\ee
where
\be
\cos\alphap =u +(1-u)\cos\psi = Q+U\cos\phi .
\ee
If secondary (antipodal) spot is present,
the corresponding flux is
\be
F_{\rm s} (\phi) = \cos\alphas [1+\ani\cos\alphas] ,
\ee
where (using $\cos\psi_{\rm s} = -\cos\psi$)
\be
\cos\alphas = u -(1-u)\cos\psi = 2u-Q-U\cos\phi .
\ee

\subsubsection{One spot}
\label{sec:ani1}

First, let us consider the simple case of black body  spots  (i.e. $\ani=0$).
The pulse profile from a single spot then has almost exactly sine shape (B02)
\be
F_p(\phi)= \cos\alphap = Q+U \cos \phi.
\ee
This immediately gives the non-zero Fourier cofficients 
   $a_0=Q$ and $a_1=U$  for pulsars in class A. 
In class B, the Fourier coefficients can be computed using 
equations~(\ref{eq:four_ecl}). The peak-to-peak amplitude
$A\equiv (F_{\max}-F_{\min})/(F_{\max}+F_{\min})$ is given by:
\be  \label{eq:ampl}
A=\left\{\begin{array}{ll}
U/Q, & {\rm class\ A,}\\
1, & {\rm class\ B}. \\
0, & {\rm class\ C}. \\
\end{array}\right.
\ee
The peak-to-peak amplitude coincides with
the ratio $|a_1|/a_0$ and is $\sqrt{2}$ larger than
sometimes quoted rms amplitude if the light curve is a pure sine profile.

The pulse profile is given by
\be \label{eq:aniflux}
F(\phi) =
Q+ \ani\left(Q^2+\frac{U^2}{2}\right) + U(1+2\ani Q) \cos\phi +
\ani\frac{U^2}{2} \cos2\phi .
\ee
One sees that anisotropy introduces no phase shift in the harmonics, and the
   pulse profile remains symmetric about $\phi=0$ and $\phi=\pi$, however
   the Fourier amplitudes are changed.
The amplitude of the  fundamental   $c_1$
increases  with  positive $\ani$ and  decreases if $\ani$ is  negative.  
The fundamental  can  even  completely  disappear  when   $\ani=-1/(2Q)$. 
 It is proportional to $U$ and therefore $c_1$ behaves close to $\sin i\sin\theta$
   (see also Fig.~\ref{fig:ampl}, left panel for class~A and right panel   for class~I).

Anisotropy $\ani\neq 0$ introduces harmonic $\cos 2\phi$ in the pulse.
The amplitude $c_2$ of this harmonic  can be quite large if $\ani$ is large 
(in absolute  value) and is proportional to $ \sin^2 i\sin^2\theta$  
(see eqs.~[\ref{eq:a2}], [\ref{eq:b2}] and lower panels in Fig.~\ref{fig:ampl}). 
The ratio of amplitudes $c_2$ and $c_1$ is given by
\be \label{eq:aniso}
%a_{1}&=&\frac{U(1+2\ani Q)}{Q+\ani(Q^2+U^2/2)}, \nonumber \\
%a_{2}&=&\frac{\ani U^2/2}{Q+\ani (Q^2+U^2/2)}, \\
   \frac{c_{2}}{c_{1}} =    \frac{a_{2}}{a_{1}}
     = \frac{\ani U/2}{1+2 \ani Q}\propto \ani \sin i \sin\theta.
%\nonumber
\ee

If $\ani\ge-1/2$,  the pulse profile has a maximum at $\phi=0$ and a minimum at $\phi=\pi$.
The corresponding pulse amplitude is
\be 
A=\frac{U(1+2\ani Q)}{Q+\ani(Q^2+U^2)} .
\ee

If $\ani<-1/2$ and the conditions 
\be\label{eq:animax} 
\cos(i+\theta)<\eta \equiv -\frac{u+1/2\ani}{1-u}<\cos(i-\theta)  
\ee
are satisfied, then the pulse has minima at both $\phi=0$ and $\phi=\pi$.
Two maxima
\be 
F_{\max}=-\frac{1}{4\ani} 
\ee
appear at phases corresponding to 
\be \label{eq:phia}
\cos\phi = \cos\phi_1 \equiv  -\frac {1+2\ani Q}{2\ani U} . 
\ee
The global minimum of the pulse profile is at $\phi=\pi$,
\be 
F_{\min}(\phi=\pi)= (Q-U)[1+\ani(Q-U)] .
\ee

If $\ani<-1/2$ and $\cos(i+\theta)> \eta$, the  pulse minimum is at $\phi=0$ and 
its maximum at $\phi=\pi$. When $\cos(i-\theta)<\eta$, then 
 the minimum is at $\phi=0$ and the maximum is at $\phi=\pi$.

In  class B, the phase $\phi=\pi$ is not visible. 
The condition for additional maxima (\ref{eq:animax}) is the same as in class A 
and these maxima  are always in the
visibility zone, $\cos\phi_1>\cos\phip$. 
  
The above analysis is also applicable to pulsars with two antipodal
spots in class~I, when only one spot is visible. 
Because class I is smaller  on the $\theta$-$i$ plane 
than class A, the existence of 
the maxima at $\phi=|\phi_1|$ (eq. [\ref{eq:phia}]) 
requires a stronger condition on $\ani$: $-1/4u<\ani<-1/2$.

\subsubsection{Two antipodal spots}
\label{sec:ani2}

First consider black-body spots  ($\ani=0$). 
In class IV, the pulse profile is flat with $a_0=2u$,
i.e. there are no pulsations.
In class II, the  pulse profile consists of a single sinusoidal pulse and
a flat plateau appearing when the antipodal spot  also
becomes visible (see eq. [\ref{eq:phis}]). 
In class III, there are two pulses with two  plateaus in between (symmetric  
 about phases $\phi=0$ and $\pi$). 
 The primary is seen at $|\phi|<\phip$, while the secondary appears 
at $\cos\phi<\cos\phis$.
The corresponding pulse amplitudes $A$ for all these classes are given by (B02)
\be  \label{eq:amplb02}
A =\left\{
\begin{array}{ll}
U/Q, & {\rm class\ I,}\\
\strut\displaystyle
\frac{U+Q-2u}{U+Q+2u}=\frac{\cos(i-\theta)-u}{\cos(i-\theta)+u}, & {\rm classes\ II,\ III,} \\
0, & {\rm class\ IV} . \\
\end{array}\right.
\ee

The anisotropy $\ani\ne0$ modifies significantly the pulse profiles.
The analysis is simplest for class IV, where two spots are always visible.
 Then
the total flux is 
\beq \label{eq:ani_four}
\bar{F}(\phi) & = & \bar{F}_{\rm p} (\phi) + \bar{F}_{\rm s} (\phi)= 
 2u+\ani\left[ 2(Q-u)^2 + 2u^2 + U^2 \right.  \nonumber \\
&+&\left.  4(Q-u)U\cos\phi+U^2\cos 2\phi  \right] 
\eeq
and the Fourier amplitudes are 
\beq \label{eq:a1ani}
   \frac{c_1}{c_0} =  \frac{a_1}{a_0} 
&=& \frac{2\ani (Q-u)U}{u+\ani [u^2 +(Q-u)^2 + U^2/2]}  , \\
\label{eq:a2ani}   \frac{c_2}{c_0} = \frac{a_2}{a_0} 
&=& \frac{\ani U^2/2}{u+ \ani [u^2 +(Q-u)^2 +U^2/2]}  .
\eeq

The dependencies 
   $c_1/c_0\propto \ani \sin 2i \sin2\theta$ and 
   $c_2/c_0 \propto \ani \sin^2 i \sin^2\theta$ are found
not only in class IV,  but also in classes~II 
and III (right panel, Fig. \ref{fig:ampl}). 
Note that the importance of the second harmonic 
\be \label{eq:a2a1ani}
    \frac{c_2}{c_1} 
=   \frac{U}{4(Q-u)}=\frac{1}{4} \tan i\tan\theta 
\ee
is independent of $\ani$ and grows with $i$ and $\theta$.

If $i+\theta\le 90\degr$,  the pulse profile has two extrema at  $\phi=0$ and $\phi=\pi$. 
The pulse peak-to-peak amplitude is
\be 
A= \frac{2(Q-u)U |\ani| }{u+\ani[u^2 + (Q-u)^2+U^2]} . 
\ee
If $i+\theta>90\degr$, there are two additional extrema 
at 
$\cos\phi=\cos\phi_2 \equiv -(Q-u)/U=-\cot i \cot\theta$ with 
\be 
F(\phi_2)=2u(1+\ani u). 
\ee
These extrema are minima if $\ani>0$ and maxima if $\ani<0$.
Here $\phi=0$ is a global maximum for $\ani>0$ and a global
minimum for $\ani<0$.
The amplitude of the pulse with  four extrema grows with $|\ani|$, 
\be 
A=\frac{(1-u)^2\ \cos^2(i-\theta) \  |\ani|}{2u+\ani [2u^2 +(1-u)^2\ \cos^2(i-\theta)  ]} .
\ee
It is maximum at the boundary of class IV when $\cos(i-\theta)=\kappa$.

\subsection{Effects of fast rotation}
\label{sec:doppler}

The effects of fast rotation are illustrated in this section with a simple 
case of black body spots ($\ani=0$, no anisotropy).
Furthermore, we assume that the spots are   always visible, 
i.e. consider classes A, I, IV. 

\subsubsection{One spot, classes A and I}
\label{sec:doppler1}

The flux from one spot is given by
\beq \label{eq:dopsim}
&& F_{\rm p} (\phi) =  \Dop^{3+\Gamma} \cos\alphap  \\
&&=  Q+ U \cos \phi - (3+\Gamma)QT \sin\phi - 
\frac{3+\Gamma}{2} UT \sin 2\phi  \nonumber \\
&&=  Q+ \frac{U}{\cos\zeta_1} \cos(\phi+\zeta_1)
+ \frac{3+\Gamma}{2} UT \cos[2(\phi+\pi/4)] , \nonumber
\eeq
which only slightly deviates from a sinusoidal shape. 
The phase shift 
\be
\tan\zeta_1= (3+\Gamma)Q\ \frac{T}{U} \approx \frac{\betaeq}{\sqrt {1-u}}(3+\Gamma) Q 
\ee
 is of the order of $\betaeq\ll1$.
The main  dependences of the Fourier amplitudes on $i$ and $\theta$ are 
similar to the anisotropic case considered in Section \ref{sec:ani1}, 
with  $c_1$ behaving close to $\sin i\sin\theta$ 
and $c_2\propto \sin^2 i\sin^2\theta$  (compare  eqs. [\ref{eq:a1}]
and [\ref{eq:b1}] as well as eqs. [\ref{eq:a2}] and [\ref{eq:b2}]). 
The ratio of harmonics is
\be \label{eq:ampldop}
\frac{c_{2}}{c_{1}} \approx \frac{(3+\Gamma)T}{2} \approx
\frac{(3+\Gamma)}{2}\  \betaeq \sqrt{1- u} \ \sin i\ \sin \theta .
\ee
We see that the
Doppler  effect  produces the first harmonic  and  introduces a phase shift
between the  fundamental  and the harmonic, which  skews the profile.
The phase shift is proportional to $\betaeq$.

We note, however,  that the  ratio  $c_{2}/c_{1}$  depends  linearly  on
$\betaeq$ (which  is a small  number)  and $\sin i\sin\theta$.
We  conclude that for a single spot  $c_{2}/c_{1}\ll1$, i.e. 
the Doppler  effect alone  cannot  introduce a strong  additional
harmonic  to the  signal (unless $\Gamma$ is large).
By contrast, an anisotropic source can make it easily 
(see eq.~[\ref{eq:aniso}]).
 In combination with anisotropy, the Doppler effect 
 significantly modifies the pulse profile.

The time delays slightly reduce the phase lag to 
$\tan \zeta_1= \betaeq[ (4+\Gamma) Q -1]/\sqrt {1-u}$ and 
produce a third harmonic with a small amplitude $a_3\propto U T^2 \propto \betaeq^2$.  
 
\subsubsection{Two spots, class IV}

The contribution of the secondary (antipodal) spot to the flux is given by 
\beq
F_{\rm s} (\phi) &=& \Dop_{\rm s}^{3+\Gamma} \cos\alphas  
= 2u-Q - U \cos \phi \nonumber \\
& +&   (3+\Gamma)(2u-Q)T \sin\phi - 
\frac{3+\Gamma}{2} UT \sin 2\phi  ,
\eeq
where $\Dop_{\rm s}\approx 1+T\sin\phi$.
The total flux is given by
\beq
F(\phi) & =&  F_{\rm p} (\phi) + F_{\rm s} (\phi) \nonumber \\
& = &   2u-  (3+\Gamma) T [ 2(Q-u) \sin\phi + U\sin 2\phi ].
\eeq
The time delays introduce the effects of the order $\betaeq^2$ which are 
ignored here.

The Doppler effect modulates the flux, so that the profile is not 
    a plateau anymore. The Fourier amplitudes $c_1$ and $c_2$ grow 
linearly with the rotational frequency and the spectral index $\Gamma$. 
The amplitude of the fundamental, 
\be \label{eq:dop_c1c0}
\frac{c_1}{c_0}= \frac{3+\Gamma}{u}(Q-u)\ T  \propto \betaeq \sin 2i \sin2\theta \\
\ee
 is maximum at the boundary of class IV where $\cos(i-\theta)=\kappa$, while 
the amplitude of the harmonic, 
\be \label{eq:dop_c2c0}
\frac{c_2}{c_0}=\frac{3+\Gamma}{2u}UT\propto\betaeq\sin^2i\sin^2\theta, \\
\ee
 is maximum when $i=\theta=90\degr$.
The dependences on $i$ and $\theta$ are similar to the 
anisotropic case from Section \ref{sec:ani2}. This 
explains the fact that when both anisotropy and fast rotation are 
present, the behaviour remains the same (see right panels in 
Fig.~\ref{fig:ampl}). The ratio of the harmonics is
\be
\frac{c_2}{c_1} = \frac{U}{2(Q-u)}=\frac{1}{2}\tan i\tan \theta .
\ee
It differs by a factor of two from the anisotropic case (\ref{eq:a2a1ani}),
and becomes large at large inclination $i$ and spot co-latitude $\theta$.

If $i+\theta\le90\degr$, the Doppler-boosted pulse from two 
antipodal spots has two extrema at 
\be 
\cos\phi= \cos\phi_+ \equiv \frac{\sqrt{8+\cot^2 i \cot^2\theta}- \cot i \cot \theta}{4} > 0 ,
\ee 
with maximum at $-\phi_+$ and minimum at $\phi_+$. 
If  $i+\theta>90\degr$, there are two additional extrema at 
\be 
\cos\phi= \cos\phi_- \equiv -\frac{\sqrt{8+\cot^2 i \cot^2\theta} + \cot i \cot \theta}{4} < 0 ,
\ee 
with  $\phi_-$ being the maximum and $-\phi_-$ being the minimum. The global 
maximum and minimum remain at $\mp\phi_+$.

\section{Summary}

\label{sec:summary} 

We have derived  an analytical approximation for 
the pulse profile produced by small spots on a neutron star surface,
 its Fourier amplitudes and phases.
The exact profile is reproduced with good accuracy by our  formulae 
even in the case of very fast rotation 
(e.g. with the error of $\sim 15$~per cent for rotational frequency
   $\nu=600$~Hz).  For slower rotation 
   of 300~Hz, the accuracy improves to $\sim 5$~per cent. 
The main advantage of the analytical formalism is that it shows the 
dependence of the pulse profile and its Fourier series on the parameters 
of the pulsar.  

The proposed formalism can be used to obtain constraints on the neutron star
parameters and position of the hot spot 
from the amplitudes of oscillations observed during X-ray bursts in some 
low-mass X-ray binaries, as well as from the X-ray pulse profiles 
of the accretion-powered and  rotation-powered pulsars. 
Our results can be further extended to study the energy dependence of the 
profile as well as the effects of the finite spot size.

\section*{Acknowledgments}

This research has been supported by 
the Academy of Finland grant 102181 and 
the Vilho, Yrj\"o and Kalle V\"ais\"al\"a Foundation.
 AMB thanks the Division of Astronomy of the University of Oulu for 
    hospitality during his visit, when this work was finished.

%#####################################################################
\appendix

\section{Exact calculation of observed flux}

\subsection{Light bending and Lorentz transformations}

The exact relation between $\alpha$ and $\psi$ in Schwarzschild geometry
(i.e. light bending) is given by \citep[e.g.][]{mtw73}
\be \label{eq:bend}
  \psi=\int_R^{\infty} \frac{dr}{r^2} \left[ \frac{1}{b^2} -
       \frac{1}{r^2}\left( 1- \frac{\rg}{r}\right)\right]^{-1/2} ,
\ee
where $b$ is impact parameter,
\be \label{eq:impact}
  b=\frac{R}{\sqrt{1-u}} \sin\alpha ,
\ee
$u=\rg/R$, $\rg=2GM/c^2$ is Schwarzschild radius; $M$ and $R$ are mass and
radius of the star. The maximum bending angle $\psi_{\max}$
corresponds to $\alpha=\pi/2$.
The visibility of the spot is defined by a condition $\cos \alpha>0$, or
alternatively by $\psi<\psi_{\max}\equiv \psi(\alpha=\pi/2)$.

As a photon emitted at angle $\alpha$ with respect to the spot normal
$\bmath{n}$ propagates to infinity, its direction changes from
$\bmath{k}_0$ near the stellar surface to $\bmath{k}$ at infinity,
so that $\cos\alpha=\bmath{k}_0\cdot\bmath{n}$
changes to $\cos\psi=\bmath{k}\cdot\bmath{n}$.
The relation between $\bmath{k}_0$ and $\bmath{k}$ may be written as
\be\label{eq:k0}
\bmath{k}_0=[ \sin\alpha\ \bmath{k} +\sin(\psi-\alpha)\ \bmath{n}]/\sin\psi.
\ee
At any moment of time, we can introduce an instantaneous non-rotating frame $x,y,z$ 
with the $y$-axis along the direction of the spot motion, 
$x$-axis along the meridian towards the equator, and 
$z$-axis along the normal  to the spot.
In this static frame, 
\be 
\bmath{k}_0=
\left( 
\frac{\sin\alpha}{\sin\psi} (\sin i\cos\theta\cos\phi -\cos i \sin\theta),
\cos\xi, 
\cos\alpha
\right) ,
\ee 
where
\be \label{eq:cosxi2}
\cos\xi=\frac{\bbeta}{\beta} \cdot \bmath{k}_0
=\frac{\sin\alpha}{\sin\psi} \frac{\bbeta}{\beta} \cdot \bmath{k}=
- \frac{\sin\alpha}{\sin\psi}\sin i\ \sin\phi\  .
\ee
In the frame comoving with the spot 
(with $y$-axis along the spot motion, $z$-axis along the local normal), 
the unit vector along the photon momentum  is 
obtained from the Lorentz transformation: 
\be \label{eq:k0prime}
\bmath{k}'_0 = \Dop
\left( \begin{array}{c}
(\sin i\cos\theta\cos\phi -\cos i \sin\theta)\sin\alpha/\sin\psi \\
\gamma (\cos\xi-\beta)\\ 
\cos\alpha
\end{array}
\right) ,
\ee 
where  $\gamma=1/\sqrt{1-\beta^2}$ and the Doppler factor 
\be \label{eq:dop}
\Dop=\frac{1}{\gamma(1-\beta\cos\xi)} .
\ee
Using equation (\ref{eq:k0prime}), we obtain
\be \label{eq:cosalphadop}
\cos\alpha'= \Dop\cos\alpha .
\ee

\subsection{Observed flux}

The observed flux from the spot 
at photon energy $E$ is
\be
\label{eq:dF_E}
  \d F_E=I_E\ \d\Omega,
\ee
where $I_E$ is the specific   intensity of radiation
at infinity and $\d\Omega$ is
the solid angle occupied by spot with area $\d S'$ on the observer's sky.
The solid angle can be expressed in terms of the impact parameter
\be
\d\Omega=b\ \d b\ \d\varphi/D^2,
\ee
where $D$ is the distance to the source and $\varphi$ is the azimuthal
angle corresponding to rotation around line of sight (vector $\bmath{k}$).
The impact parameter $b$ depends on $\psi$ only, but not on $\varphi$.

Using equation~(\ref{eq:impact}) and
the facts that $\d S=R^2\ \d\cos\psi\  \d\varphi$ and
$\d S'\ \cos\alpha'= \d S\ \cos\alpha$
(since the spot area projected on to the plane perpendicular
to the photon propagation direction, i.e. a photon beam cross-section,
is Lorentz invariant), one gets
\be\label{eq:omega}
 \d\Omega=\frac{\d S' \cos \alpha'}{D^2} \frac{1}{1-u} \frac{\d\cos\alpha}{\d\cos\psi}.
\ee
In the limit of weak gravity $u\ll 1$, this gives
the usual formula $\d\Omega=\d S' \cos \alpha'/D^2$.

The combined effect of the gravitational redshift and Doppler effect
results in the following  relation between the monochromatic
observed and local intensities
\citep[see e.g.][]{mtw73,rl79}:
\be
I_{E} = \left (\frac{E}{E'}\right )^3 I'_{E '} (\alpha')
\ee
where $E/E'=\Dop \sqrt{1-u}$.
Here $I'_{E'}(\alpha')$ is the intensity computed in the frame comoving with
the spot.
For the bolometric intensity, one gets
\be
I= \left (\Dop \sqrt{1-u} \right )^4 I'(\alpha') .
\ee
If the radiation spectrum can be represented by
a power-law $I'_{E'}(\alpha') \propto E '^{-(\Gamma-1)}$
with a photon spectral index
$\Gamma$ which does not depend on the angle $\alpha'$ then
\be \label{eq:int_trans}
I'_{E'}(\alpha') = I'_{E}(\alpha')
\left( \Dop \sqrt{1-u} \right)^{\Gamma-1} .
\ee
This approximation is equivalent to the assumption of a weak
energy dependence of the angular distribution \citep[see][]{vp04}.

The observed spectral flux (eq.~\ref{eq:dF_E}) now reads
\be \label{eq:fluxspot}
\d F_{E}=(1-u)^{1/2} \Dop^{4} I'_{E'}(\alpha') \cos\alpha
\frac{\d \cos\alpha}{\d\cos\psi}
 \frac{\d S'}{D^2} ,
\ee
where we have used the aberration formula (\ref{eq:cosalphadop}).
Substituting equation (\ref{eq:int_trans}), we recover
equation (\ref{eq:fluxplaw}).
 
The bolometric flux is given by:
\be
\d F= (1-u)\ \Dop^5 \
I'(\alpha')  \cos\alpha \frac{\d\cos\alpha}{\d\cos\psi} \frac{\d S'}{D^2} .
\ee
Thus, the flux from a rapidly rotating star differs by a factor  $\Dop^5$
from that from a slowly rotating star \citep{pg03}.  Two powers of $\Dop$ come
from the solid angle transformation, one from the energy, one from the
photon arrival time contraction,
and the fifth from the change in the projected  area due to  aberration.
Aberration may also change the specific intensity since it has to be computed
for angle $\alpha'$ in the comoving frame.
  
\subsection{Time delays} 
  
Finally, we write down here the formula describing the time delays.
The delay is caused by different travel times of emitted
photons to the observer, depending on the position of the emitting spot.
A photon following the trajectory with an impact parameter $b$
is lagging the photon with $b=0$ by \citep{pfc83}:
\be \label{eq:delay}
c\Delta t(b)=  \int_R^{\infty} \frac{\d r}{1- \rg/r}
\left\{ \left[ 1-  \frac{b^2}{r^2}  \left( 1- \frac{\rg}{r} \right)
\right] ^{-1/2}  -1 \right\} .
\ee
For a given pulsar phase $\phi$, we compute angle $\psi$, then
we find the corresponding emitted $\alpha$ and the impact parameter
using  formulae (\ref{eq:bend}) and (\ref{eq:impact}), and
  compute the corresponding  delays $\Delta t(b)$
with equation  (\ref{eq:delay}).
We then construct a one-to-one
correspondence between the pulsar phase $\phi$ and
the photon arrival phase to the observer $\phiobs=\phi+ \Delta\phi$,
with the phase delays
\be \label{eq:deltaphi}
\Delta \phi(\phi) =2\pi\nu \Delta t[b(\phi)] .
\ee
For analytical work we can also use the relation $\phi=\phiobs-\Delta\phi(\phi)
\approx \phiobs-\Delta\phi(\phiobs)$.
 
\label{lastpage}

\begin{thebibliography}{99}
\bibitem[\protect\citeauthoryear{Beloborodov}{2002}]{b02}
  Beloborodov A. M., 2002, ApJ, 566, L85 (B02)
\bibitem[\protect\citeauthoryear{Gierli\'nski \& Poutanen}{2005}]{gp05}
Gierli\'nski M., Poutanen J., 2005,  MNRAS, 359, 1261
\bibitem[\protect\citeauthoryear{Gierli\'nski, Done \& Barret}{Gierli\'nski et al.}{2002}]{gdb02}
  Gierli\'nski M., Done C.,  Barret D., 2002, MNRAS, 331, 141
\bibitem[\protect\citeauthoryear{Leahy \& Li}{1995}]{ll95}
Leahy D. A.,  Li L., 1995, MNRAS, 277, 1177
\bibitem[\protect\citeauthoryear{Miller \& Lamb}{1998}]{ml98}
  Miller M. C., Lamb F. K., 1998, ApJ, 499, L37
\bibitem[\protect\citeauthoryear{Misner, Thorn \& Wheeler}{Misner et al.}{1973}]{mtw73}
  Misner C. W., Thorn K. S.,  Wheeler J. A., 1973,
        Gravitation. Freeman, San Francisco
\bibitem[\protect\citeauthoryear{Muno et al.}{2002}]{moc02}
        Muno M. P., \"Ozel F., Chakrabarty D., 2002, ApJ, 581, 550
\bibitem[\protect\citeauthoryear{Pechenick, Ftaclas \& Cohen}{Pechenick et al.}{1983}]{pfc83}
  Pechenick K. R., Ftaclas C., Cohen  J. M., 1983, ApJ, 274, 846
\bibitem[\protect\citeauthoryear{Poutanen}{2006}]{p06}
Poutanen J., 2006, Adv. Space Res., in press [arXiv:astro-ph/0510038]
\bibitem[\protect\citeauthoryear{Poutanen \& Gierli\'nski}{2003}]{pg03}
 Poutanen J.,  Gierli\'nski M., 2003, MNRAS, 343, 1301
\bibitem[\protect\citeauthoryear{Rybicki  \& Lightman}{1979}]{rl79}
Rybicki G. B.,  Lightman A. P., 1979,   Radiative Processes in Astrophysics.
      Wiley-Interscience, New York
\bibitem[\protect\citeauthoryear{Strohmayer \& Bildsten}{2006}]{sb06}
  Strohmayer T., Bildsten L., 2006, in Lewin W. H. G,
  van der Klis M., eds, Compact Stellar X-Ray Sources. Cambridge University Press,
  Cambridge, p. 113 (astro-ph/0301544)
  \bibitem[\protect\citeauthoryear{Strohmayer et al.}{2003}]{s03}
  Strohmayer T. E., Markwardt C. B., Swank J. H., in 't Zand J., 2003, ApJ, 596, L67
  \bibitem[\protect\citeauthoryear{Viironen \& Poutanen}{2004}]{vp04}
  Viironen K.,  Poutanen J., 2004, A\&A, 426, 985 
\bibitem[\protect\citeauthoryear{Weinberg, Miller \& Lamb}{Weinberg et al.}{2001}]{wml01}
  Weinberg N., Miller M. C., Lamb D. Q., 2001, ApJ, 546, 1098
\bibitem[\protect\citeauthoryear{Wijnands}{2006}]{w05}
Wijnands R., 2006, in Lowry J. A., ed, Trends in Pulsar Research. 
Nova Science Publishers, New York, in press (astro-ph/0501264)
\bibitem[\protect\citeauthoryear{Zavlin, Shibanov \& Pavlov}{Zavlin et al.}{1995}]{zsp95}
  Zavlin V. E., Shibanov Yu. A., Pavlov G. G.., 1995, Astron. Let., 21, 149
\end{thebibliography}
\end{document}